\documentclass[structabstract]{aa}
\usepackage{epsfig,amsmath,amssymb}

\def\msol{M_\odot}
\def\rsol{R_\odot}

\def\lsol{L_\odot}

\def\simgr{\,\hbox{\hbox{$ > $}\kern -0.8em \lower 1.0ex\hbox{$\sim$}}\,}
\def\simle{\,\hbox{\hbox{$ < $}\kern -0.8em \lower 1.0ex\hbox{$\sim$}}\,}
\def\beq{\begin{equation}}
\def\eeq{\end{equation}}

\begin{document}

\title{Accuracy of stellar parameters of exoplanet-host stars determined from asteroseismology}

\author{C. Mulet-Marquis\inst{1}, I. Baraffe\inst{1}, S.Aigrain\inst{2}, F.Pont\inst{2}}

\offprints{C. Mulet-Marquis, S.Aigrain}

\institute{C.R.A.L, CNRS, UMR5574
 \'{E}cole normale sup\'erieure, 46 all\'ee d'Italie, 69007 Lyon, France (cedric.mulet-marquis, ibaraffe @ens-lyon.fr)
\and
School of Physics, University of Exeter, Stocker Road, Exeter, EX4 4QL, UK (suz, fpont @astro.ex.ac.uk)}

\date{Received /Accepted}
\titlerunning{Accuracy of stellar parameters determination}
\authorrunning{Mulet-Marquis {\it et al.} }
\abstract
{In the context of the space-based mission CoRoT, devoted to 
asteroseismology and search for planet transits, we analyse the accuracy of  fundamental stellar parameters  (mass, radius, luminosity) that can be obtained from asteroseismological data.} 
{Our work is motivated by the large uncertainties on planetary radius determination
of transiting planets which are mainly due to uncertainties on the stellar parameters. 
Our goal is to analyse uncertainties of fundamental stellar parameters for a given accuracy of oscillation frequency determination. }
{We generate grids of equilibrium models of stars and compute their pulsation spectra  based on a linear nonadiabatic stability analysis. Using differents methods of comparison of oscillation mode spectra, we derive uncertainties on fundamental stellar parameters and analyse the effect of varying the  number of considered modes. } 
{The limits obtained depend strongly on the  adapted method to compare spectra. We find a degeneracy in the stellar parameter solutions, up to a few \% in mass (from less than 1\% to more than 7\% depending on the method used and the number of considered modes), luminosity (from 2\% to more than 10\%) or radius (from less than 1\% to 3\%), for a given pulsation spectrum. } 
{}
\keywords{Asteroseismology} 
\maketitle
\section{Introduction}
The successful launching of CoRoT produces data for transiting exoplanets with unprecedented accuracy compared to ground-based observations, providing a measure of planetary mass and radius and thus information on the mean density and bulk composition of exoplanets. Unfortunately, a remaining source of uncertainty on the planetary parameters (mass and radius) is due to uncertainties on the stellar parameters. 

The precise determination of gravity by spectroscopy is difficult, and particularly for F stars. The determination of the mass of the star is the main uncertainty on the planet's size. This uncertainty on the mass is generally about 10\% ({\it i.e} 3\% on the radius and 10\% on the density). This may be of importance to disantangle a massive planet
from a brown dwarf, as in the case of the recently discovered "super-Jupiter"
CoRoT-Exo-3b (Leconte et al. 2009). It may also prevent distinguishing an ocean planet from an earth-like planet (Grasset et al. 2009).

Our goal is to examine the accuracy on the latter parameters , namely mass, luminosity and radius, that could be obtained using asteroseismology with the expected accuracy on oscillation frequencies of CoRoT.  We explore the space of stellar parameters, in terms of mass, effective temperature, luminosity, metallicity and mixing length parameter and we analyse the sensitivity of predicted spectrum of oscillation frequencies to these parameters. 
Adopting various levels of stellar noise and white noise, we analyse the frequency uncertainty for oscillation modes of a given amplitude and lifetime. This analysis is based on  the performances of CoRoT in the Asteroseismology and Planet Finder channels described in Auvergne et al. (2009).

\section{Method}

\subsection{Stars studied}

Among all the transiting systems discovered so far, the large difference between the fundamental parameters of XO-3 given by Johns-Krull et al. (2008) (M=1.41 $\pm$ 0.08 $\msol$ ; R=2.13 $\pm$ 0.21 $\rsol$) and by Winn et al. (2008) (M=1.213 $\pm$ 0.066 $\msol$ ; R=1.377 $\pm$ 0.083 $\rsol$) was a strong motivation to examine how accurately fundamental parameters of a star (mass, radius, luminosity) can be 
constrained by asteroseismology.
In this preliminary study, we first analyse stars with characteristics close to the ones of XO-3 (M $\sim$ 1.4 $\msol$ , R $\sim$ 1.5 $\rsol$, Teff $\sim$ 6400 K). We focus on modes with low order $l$ ($0 \leq l \leq 3$) and radial order $n$ between 5 and 20.

\subsection{Frequency computation}
The pulsation calculations are performed with a nonradial code originally developed by Lee (1985) and based on a linear non-adiabatic stability analysis.
The equations are linearised around hydrostatic equilibrium, and eigenfunctions are expressed with spherical harmonics $Y_{lm}$. The eigenfrequencies are defined by

\beq
\sigma= \sigma_r+i\sigma_i, 
\eeq

with $\sigma_r$ the oscillation frequency and  $\sigma_i$ the damping rate (if positive) or growth rate (if negative) (see details in Mulet-Marquis et al. 2007, and references therein).
The system of equations is solved with a Henyey-type relaxation method. 
Stellar structure models are calculated using the Livermore opacities (Iglesias, Rogers, 1996). The number of gridpoints of the equilibrium models (4000 gridpoints) is high enough for the uncertainty on the frequency determination by the linear code to
be less than the accuracy expected from CoRoT for a pulsation mode with an infinite lifetime and observed for 150 days ({\it i.e} 0.1 $\mu$Hz).

\subsection{Accuracy of frequency determination \label{detfreq}}

The photometric performance of CoRoT, described in Auvergne et al. (2009),
allows for the detection and characterisation of Sun-like oscillations
of bright ($6<V<9$) stars in the asteroseismology (AS) channel, and
larger amplitude red giant pulsations on fainter ($V>11.5$) stars in
the planet finding (PF) channel.

The precision to which mode frequencies can be determined depends on
the mode lifetime $\Gamma$, the duration $T$ of the dataset, the mode
amplitude $A$ and the background, power density $B$ (including stellar
and white noise) at the frequency of the mode. In the case of both
Sun-like and red giant oscillations, mode lifetimes are of the order
of a few days (Fletcher et al. 2006), much shorter than a typical CoRoT run (21
or 150 days). We therefore use the formula given by Libbrecht (1992) for
the case where $\Gamma \ll T$ to estimate the frequency uncertainty $\sigma_{\nu}$ as:
\begin{equation}
\label{eq:sigma}
  \sigma_{\nu}^2 = f(\beta) \frac{\Gamma}{4 \pi T},
\end{equation}
where $\beta = B/A$ is the inverse signal-to-noise ratio, and
\begin{displaymath}
  f(\beta) = \left( 1 + \beta \right)^{1/2}
     \left[ \left(1 + \beta \right)^{1/2} + \beta^{1/2} \right]^3
\end{displaymath}

As is common practice in asteroseismology (see e.g. Kallinger et al. 2009 and
references therein), we model $B$ as a sum of broken power-laws of the
form
\begin{equation}
\label{eq:background}
  B(\nu) = \sum_i \frac{a_i}{1 + (2\pi b_i\nu)^{c_i}} + d
\end{equation}
where each term in the sum corresponds to the stellar variability
associated with a particular type of surface structure (typically
active regions and granulation, with in some cases a super-granulation
component), $a_i$ is the amplitude, $b_i$ the characteristic
time-scale, and $c_i$ the slope of term $i$, and $d$ represents the
white noise. 

\begin{table*}[h]
  \centering
  \begin{tabular}{lcccccccccccc}
    \hline\hline
    Case & $V$ & $A$ & $d$ & $a_1$ & $b_1$ & $c_1$ & $a_2$ & $b_2$ & $c_2$ & $a_3$ & $b_3$ & $c_3$ \\
    \hline
    active Sun & 6 & 10 & 0.65 & $2.0\cdot10^4$ & $2.0\cdot10^6$& 4.35 & 5.51 & 8205 & 3.5 & 0.19 & 341 & 1.4 \\
    $10 \times$ active Sun & 8 & 10 & 1.95 & $2.0\cdot10^5$ & $2.0\cdot10^6$& 4.35 & 55.1 & 8205 & 3.5 & 1.9 & 341 & 1.4 \\
    red giant & 12 & 1000 & 6.8 & $3.5\cdot10^5$ & $1.2\cdot10^5$  & 4.0 & 55.1 & 6367 & 3 & -- & -- & -- \\
    \hline
  \end{tabular}
  \caption{Parameters of the background models used. The photon noise level $d$ is given in ppm$/\sqrt{\mu{\rm Hz}}$, while the background amplitudes $a_i$ are given in ppm$^2/\mu$Hz, and the time-scales $b_i$ in $10^6$\,s. Three components were used for the Sun-like cases but 2 only for the red giant case.}
  \label{tab:CoRoTpar}
\end{table*}

\begin{figure}[h]
  \psfig{file=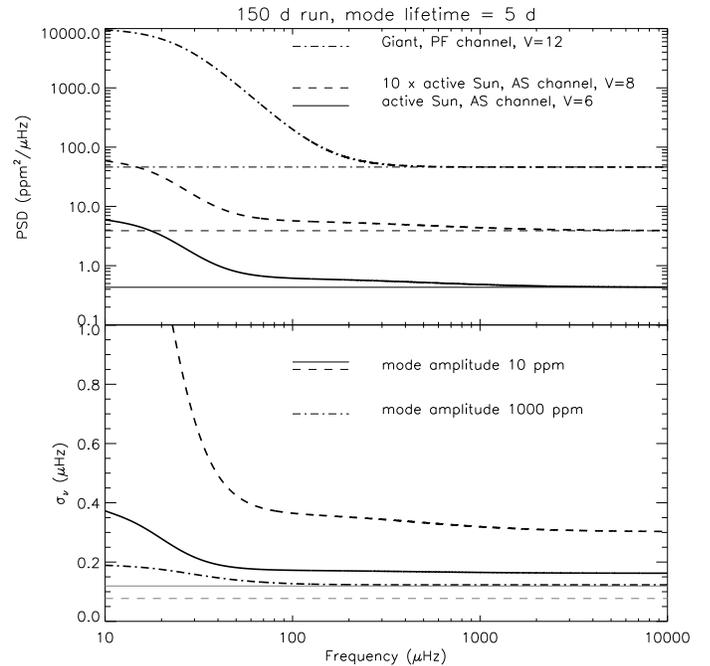,width=\linewidth}
  \caption{Top panel: Typical background power spectrum density
    $B(\nu)$ (see text for details) for a bright Sun-like star in the
    CoRoT AS channel (solid curve), a fainter and more active star in
    the same channel (dashed curve), and a red giant at the bright end
    of the PF channel (dash-dot curve). The thinner horizontal lines
    show the white noise contribution in each case. Bottom panel:
    resulting frequency uncertainties for modes with a lifetime of 5 days
    observed for 150 days, with amplitudes 10 ppm (AS channel) and
    1000 ppm (PF channel). Also shown are the minimum uncertainty
    arising from the finite mode lifetime (horizontal solid grey
    line) and the minimum uncertainty arising from the finite run
    duration for an infinite mode lifetime (Nyquist sampling,
    horizontal dashed grey line).}
  \label{precisionCoRoT}
\end{figure}

As a proxy for the background signal produced by a typical star in
which one might search for Sun-like oscillations, we fit a 3-component
background (following Eq.~\ref{eq:background}) to total solar
irradiance data obtained by the VIRGO/PMO6 radiometer on the SoHO
satellite during a 5-month period close to the last solar activity
cycle maximum (for details see Aigrain et al. 2004). We then added white
noise at the level expected for a bright ($V=6$) star in the CoRoT
asteroseismology (AS) channel (all white noise values from Auvergne et al. 2009). The resulting background power density spectrum is
shown as the thick solid line in the top panel of
Fig.~\ref{precisionCoRoT}, over the frequency range where oscillations
are expected. We also constructed a less favourable case by scaling up
the solar background by a factor of 10 (this is not unreasonable, as the
Sun is generally thought to be a relatively quiet star) and adding
white noise corresponding to a $V=8$ star observed by CoRoT (thick
dashed line in Fig.~\ref{precisionCoRoT}). Finally we also constructed
a red-giant case by visually estimating the power density spectrum
parameters for such an object from Fig.~1 of Kallinger et al. (2009) and adding
white noise corresponding to a $V=12$ star observed by CoRoT (thick
dash-dot line in Fig.~\ref{precisionCoRoT}). The adopted background power
density parameters are summarised in Table~\ref{tab:CoRoTpar}.

We then computed frequency uncertainties as a function of mode
frequency, using Eq.~(\ref{eq:sigma}), for mode amplitudes of
$A=10$\,ppm (sun-like oscillation in AS channel) or $A=1000$\,ppm (red
giant oscillations in PF channel). The results are shown in the bottom
panel of Fig.~\ref{precisionCoRoT}.  In both cases, we assumed a mode
lifetime of $\Gamma=5$\,days and a run duration of $T=150$\,days
(CoRoT long run).

\section{Analysis of theoretical oscillation spectra\label{resultats}}

We first generate a coarse grid of equilibrium stellar models with various values of mass ($ 1.35 \leq \frac{M}{\msol} \leq 1.45$, $ \frac{\delta M}{\msol} = 0.01$), luminosity ($ 0.55 \leq log(\frac{L}{\lsol}) \leq 0.65$, $ \delta (log(\frac{L }{\lsol})) = 0.01$) and effective temperature ($ 6300 K \leq T_{eff} \leq 6500K$, $ \delta T_{eff}= 25 K$). 
A refined grid is then generated around the stellar parameters which predict
close pulsation spectra, according to the results obtained from the coarse grid.
 On this refined grid ($ \delta (log(\frac{L }{\lsol})) = 0.005$ , $ \delta T_{eff}= 10 K$, but constant mass because of time computation), we also vary the mixing-length parameter ($ 1.6 \leq \alpha_{MLT} \leq 1.8$, $ \delta \, \alpha_{MLT}= 0.05 $), the metallicity and helium content $Y_{He}$ ($ 0.26 \leq Y_{He} \leq 0.28 $, $ \delta Y_{He}= 0.005$) of the equilibrium model. For this preliminary study, the metallicity $Z$ is varied between $Z=4.10^{-3}$ and $Z=0.02$. In a forthcoming study, 
we plan to include a finer grid for the stellar mass and models with over-solar metallicities.

We use two methods for the comparison between  pulsation spectra computed from
 two different sets of stellar parameters: 
\begin{itemize}
\item the maximum difference of the frequencies $ \delta \nu $ of the spectra for given values of the radial order $n$ and of the degree $l$ of the mode,
\item the maximum difference of the large separation $\Delta \nu = \nu_{n,l} - \nu_{n-1,l} $ or small separation $\delta \nu_{02} = \nu_{n,0} - \nu_{n-1,2} $, (Kjeldsen et al. 2008).
\end{itemize}

A comparison of these two methods is given below.

\subsection{ {\bf Effect of the mixing length parameter $\alpha_{MLT}$ and of the metallicity $Z$} }

\begin{figure}
\psfig{file=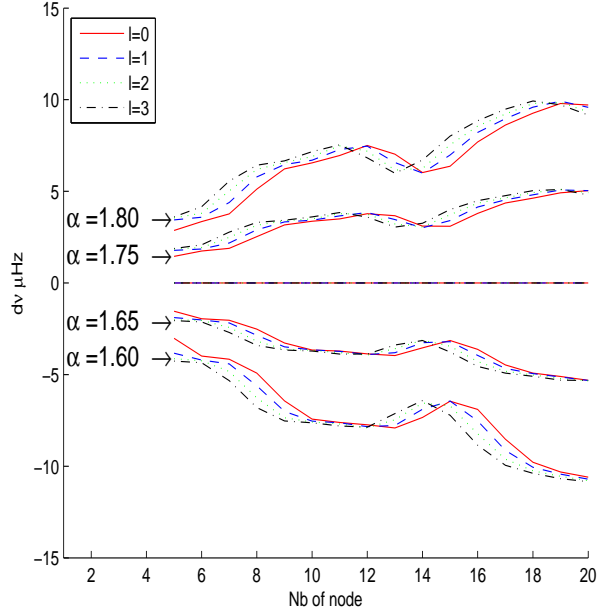,height=90mm,width=90mm} 
\caption{ Frequency difference $ d\nu $ ($\mu$Hz) as a function
of the radial order $n$, for different values of $\alpha_{MLT}$ compared to the reference value $\alpha_{MLT} = 1.7$. The solid red/dashed blue/dotted green/dash-dotted black curves correspond to degree $l$ of the mode equal to 0/1/2/3 respectively.
The horizontal line is a guide line which corresponds
to a frequency difference $d\nu$=0.}
\label{dnualpha}
\end{figure}

\begin{figure}
\psfig{file=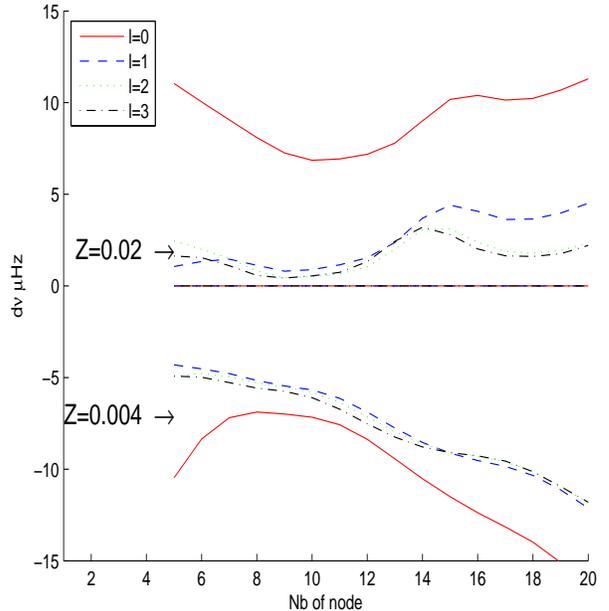,height=90mm,width=90mm} 
\caption{ Frequency difference $ d\nu $ ($\mu$Hz) as a function
of the radial order $n$, for different values of metallicity $Z$ compared to the reference value $Z = 0.01$. The solid red/dashed blue/dotted green/dash-dotted black curves correspond to degree $l$ of the mode equal to 0/1/2/3 respectively.
The horizontal line is a guide line which corresponds
to a frequency difference $d\nu$=0.}
\label{dnuz}
\end{figure}

 We examine the effect of the mixing length parameter $\alpha_{MLT}$  on the spectra in the following way : we fix $M$, $L$, $T_{eff}$, $Y_{He}$ and $Z$ and compute, for a given mode $(n,l)$, the difference $d\nu$ between the frequency for a value of $\alpha_{MLT}$ and the frequency for a reference value of $\alpha_{MLT}$ (in this case $\alpha_{MLT, ref}=1.7$). The results are shown in Figure \ref{dnualpha}.
We performed a similar exercise to estimate the effect of the metallicity $Z$. The results are displayed in Fig. \ref{dnuz} using a reference value for the metallicity $Z_{ref}$=0.01. Our preliminary tests show that a variation of $\alpha_{MLT}$ or of $Z$, within reasonable values, can yield a variation of the frequency of a given mode of several $\mu$Hz. These effects need to be investigated in a more systematic study, which we plan for a forthcoming work.

\subsection{Comparison between abolute values of frequencies \label{diffreq}}

\begin{figure}
\psfig{file=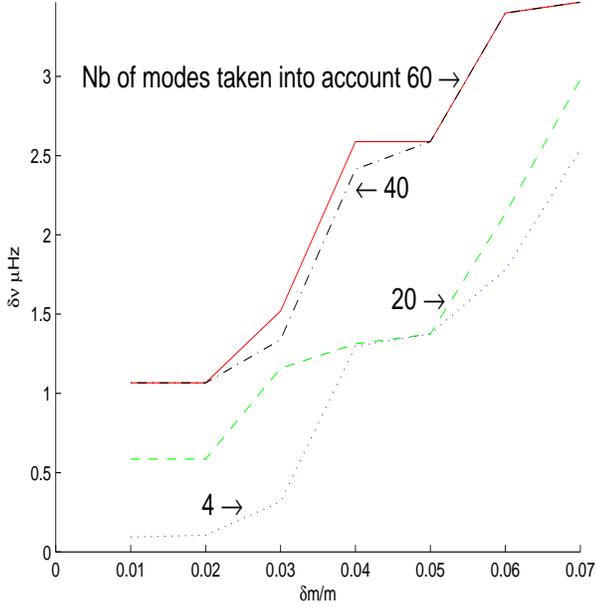,height=90mm,width=90mm} 
\caption{Spectra difference $ \delta \nu $ ($\mu$Hz) as a function
of the relative mass difference $ \delta M / M$ for the coarse grid of stellar
models (see \S \ref{diffreq}). The different curves correspond to different numbers of considered modes,
as indicated near the curves.}
\label{deltanudeltam}
\end{figure}

\begin{figure}
\psfig{file=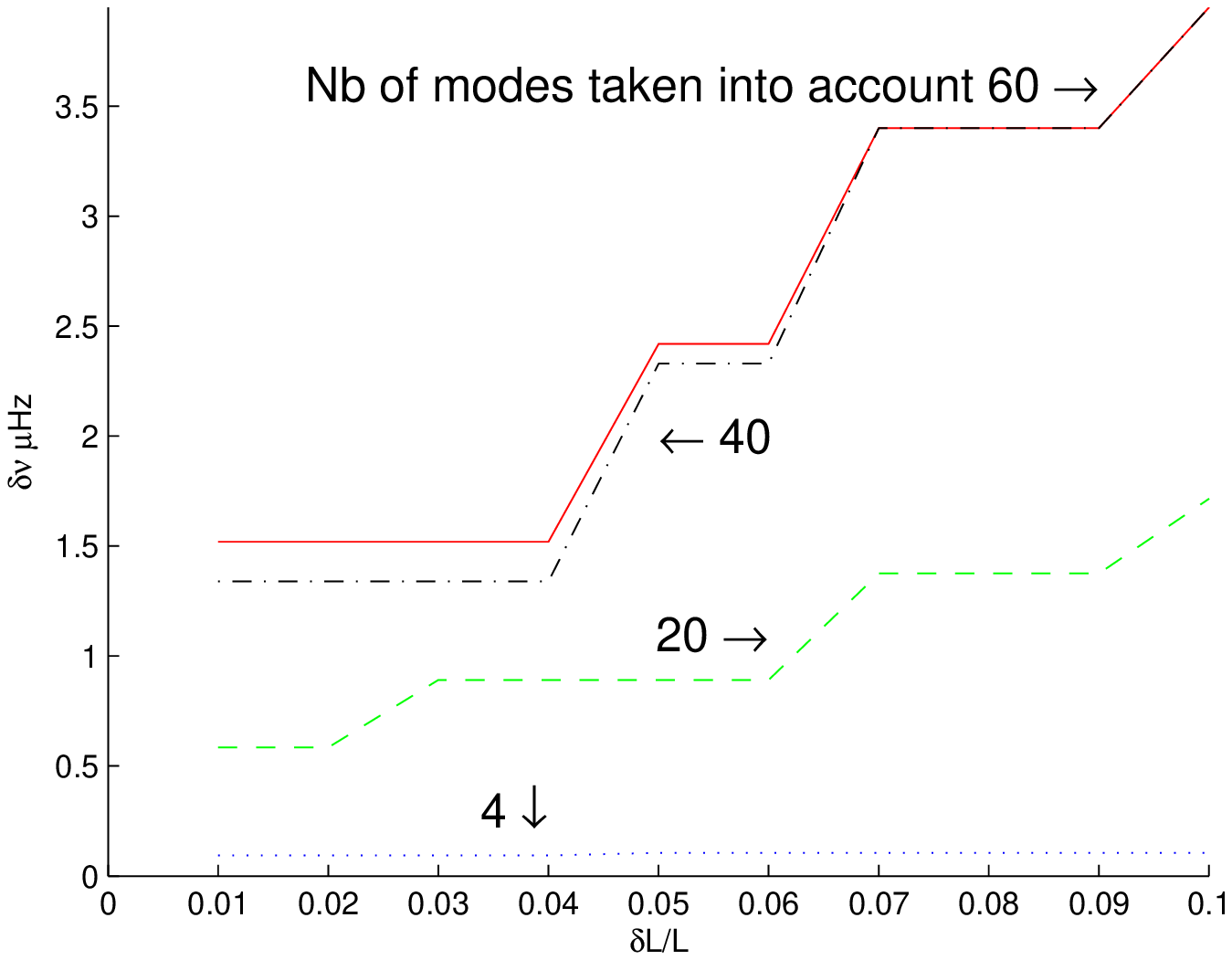,height=90mm,width=90mm} 
\caption{Spectra difference $ \delta \nu $ ($\mu$Hz) as a function
of the relative  luminosity difference $ \delta L / L$ for the coarse grid of stellar
models (see \S \ref{diffreq}). The different curves correspond to different numbers of considered modes,
as indicated near the curves.}
\label{deltanudeltal}
\end{figure}

\begin{figure}
\psfig{file=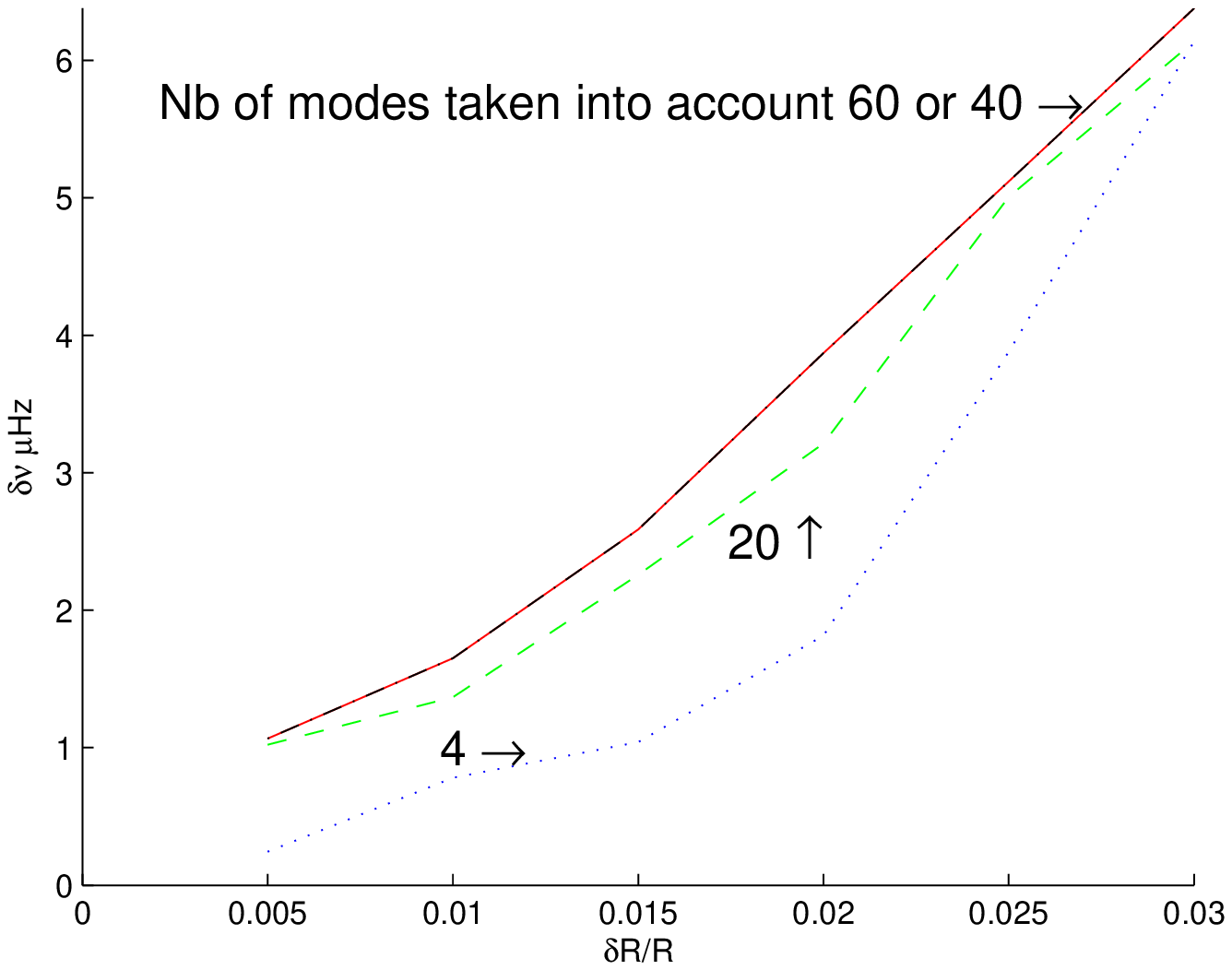,height=90mm,width=90mm} 
\caption{Spectra difference $ \delta \nu $ ($\mu$Hz) as a function
of the relative radius difference $ \delta R / R$ for the coarse grid of stellar
models (see \S \ref{diffreq}). The different curves correspond to different numbers of considered modes,
as indicated near the curves.}
\label{deltanudeltar}
\end{figure}

For a given relative mass difference $ \delta M / M$, the minimum value  $ \delta \nu_{min}$ of $ \delta \nu $ is determined in the coarse grid. The value of $ \delta \nu_{min}$ depends on the number of considered modes. 
We vary this number from 4 (one mode for each value of $0 \leq l \leq 3$) to 60 (15 modes for each value of $0 \leq l \leq 3$). The variation of $ \delta \nu_{min}$ with $ \delta M / M$, for different numbers of considered modes, 
is shown in figure \ref{deltanudeltam}.
The same procedure is applied for a given luminosity difference $ \delta L / L$ or a given radius difference $ \delta R / R$. The results are shown in figure \ref{deltanudeltal} and figure \ref{deltanudeltar} respectively.

\begin{figure}
\psfig{file=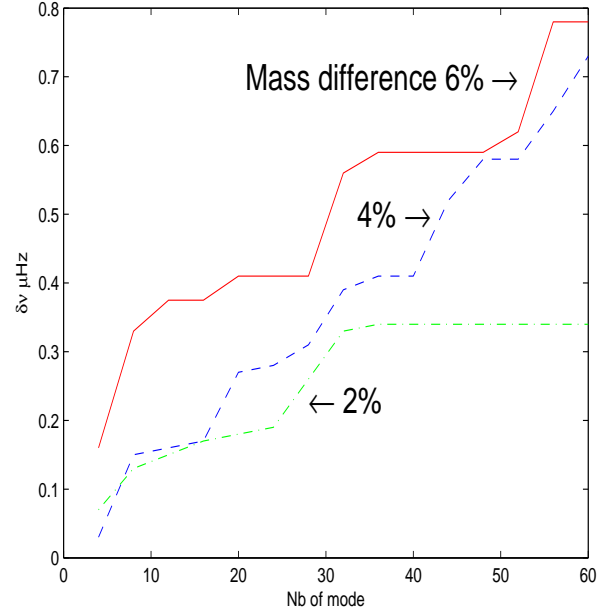,height=90mm,width=90mm} 
\caption{ Spectra difference $ \delta \nu $ (see \S \ref{diffreq}) as a function
of the number of modes for the fine grid of stellar models. The different curves correspond to different
values of the relative mass difference $ \delta M / M$.}
\label{deltanunbmode}
\end{figure}

As expected, the larger the mass/luminosity/radius difference, the larger the spectra difference.
Similarly,  the larger the number of considered modes, 
the larger the spectra difference. These trends are the same on the refined grid, but the values for $ \delta \nu_{min}$ are much smaller. If we focus on the mass difference, for instance for 40 modes, and $ \delta M / M$ = 0.06, $ \delta \nu_{min}$ decreases from 3.4 $\mu$Hz on the coarse grid to 0.6 $\mu$Hz on the fine grid, as illustrated in Fig. \ref{deltanunbmode}. The effect is similar for the luminosity difference (for 40 modes, and $ \delta L / L$ = 0.06, $ \delta \nu_{min}$ decreases from 2.4 $\mu$Hz on the coarse grid to 0.5 $\mu$Hz on the fine grid) or the radius difference (for 40 modes, and $ \delta R / R$ = 0.03, $ \delta \nu_{min}$ decreases from 6.4 $\mu$Hz on the coarse grid to 1.0 $\mu$Hz on the fine grid).
Since we cannot freely refine the grid because of computing time, the values obtained for $ \delta \nu_{min}$ on the refined grid may be overestimated : we cannot exclude that a finer grid gives smaller values of $ \delta \nu_{min}$.

\subsection{Large and small separation \label{separation}}

\begin{figure}
\psfig{file=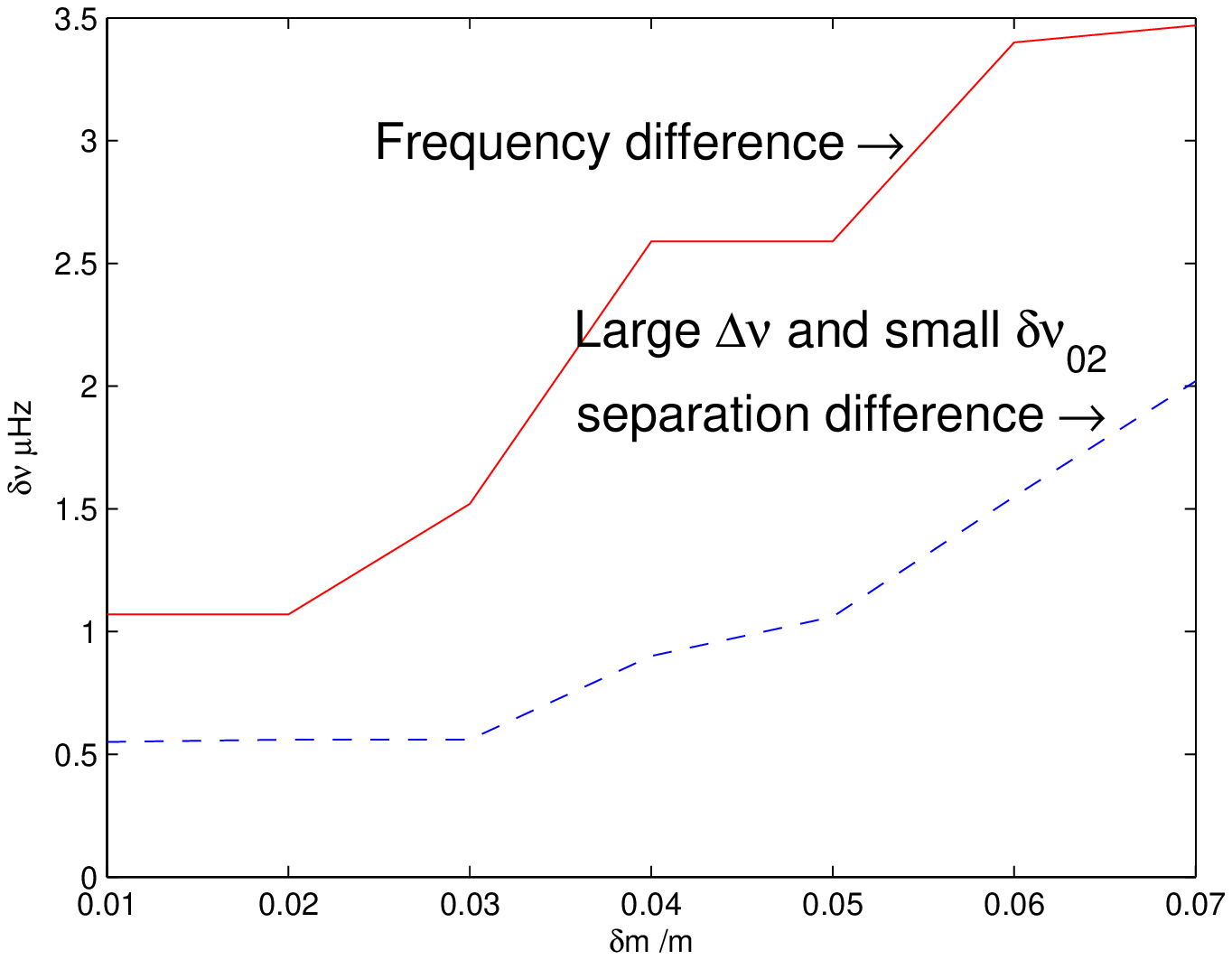,height=90mm,width=90mm} 
\caption{Comparison  between two methods measuring  spectra differences
for the coarse grid of models and with 60 considered modes. 
Red/solid line : frequency difference (\S \ref{diffreq}), blue/dashed line : large or small separation difference (\S \ref{separation}).
}
\label{petitesepa}
\end{figure}

Many asteroseismological studies do not directly compare the pulsation frequencies of two stars but rather compare the large and small separations. For this second method, we define the difference $ \delta_{sls}$ between two spectra as the maximum value of the large separation or of the small separation (the highest value is retained). 
The results for the coarse grid, for the mass difference $ \delta M / M$ with 60 considered modes,
are shown in Fig.  \ref{petitesepa}. $ \delta_{sls}$ is about a factor of two lower than $\delta \nu $. For the luminosity difference, $ \delta_{sls}$ is also about a factor of two smaller than $\delta \nu $. For the radius difference, $ \delta_{sls}$ is about a factor of three lower than $\delta \nu $.

\begin{figure}
\psfig{file=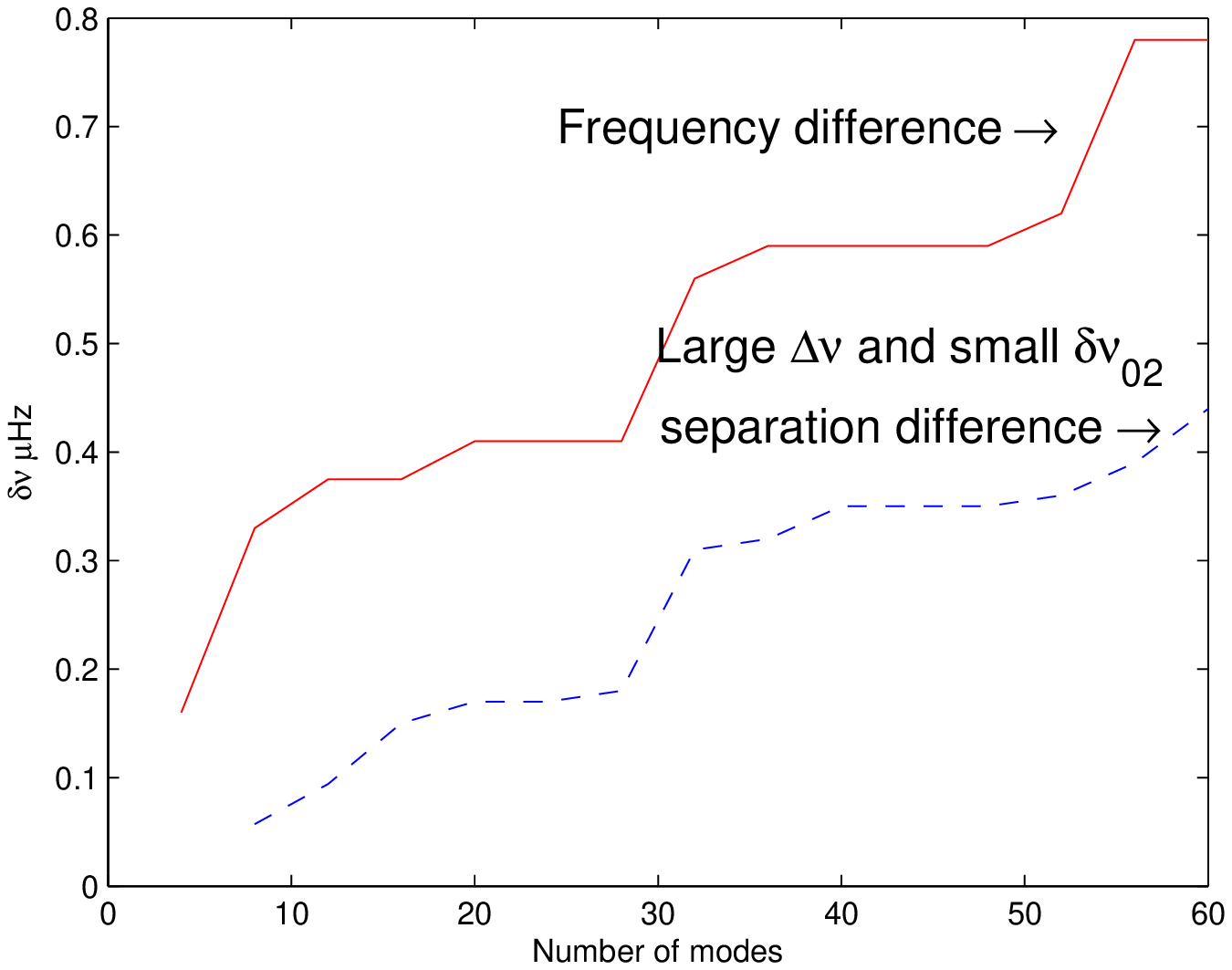,height=90mm,width=90mm} 
\caption{Comparison between two methods measuring  spectra differences
for the fine grid of models and  for $ \delta M / M$ = 0.06. Red/solid line : frequency difference (\S \ref{diffreq}), blue/dashed line : large or small separation difference (\S \ref{separation}).
}
\label{petitesepafin}
\end{figure}

As in paragraph \ref{diffreq}, the same behaviour is found when using the fine grid for a given $ \delta M / M$, $ \delta L / L$ or $ \delta R / R$. The results corresponding to a given $ \delta M / M$ are shown in Fig. \ref{petitesepafin}.


We also examined a third method, where the difference $<\delta_{sls}>$ between two spectra is defined as the difference between the mean large separation or the mean small separation (the highest value is retained). We did not find this method
relevant in the present context ($<\delta_{sls}>$ is found to be negligible for small values of $ \delta M / M$, $ \delta L / L$ or $ \delta R / R$ in the grids studied).

\section {Discussion and conclusion}


The results obtained in \S \ref{resultats} enable us to derive the accuracy of the stellar parameters mass M, luminosity L and radius R determined from asteroseismological data. The accuracy for one parameter is obtained letting all the other parameters remain free. According to \S \ref{detfreq}, we adopt the values of 0.2 $\mu$Hz and 0.3 $\mu$Hz 
 respectively for the precision of frequency determination  based on the CoRoT performances. The smaller value  is for a red giant in the PF channel and the larger value corresponds to the accuracy obtained for a Sun-like star in the AS channel
(see Fig. \ref{precisionCoRoT}).

 The results are given in Tables \ref{tabres2}, \ref{tabres2L} and \ref{tabres2R} for an accuracy of the frequency determination of 0.2 $\mu$Hz, and in tables \ref{tabres1}, \ref{tabres1L} and  \ref{tabres1R} for an accuracy of the frequency determination of 0.3 $\mu$Hz.
The most stringent constraint is provided by the method using $\delta \nu $. 
For Sun-like stars as observed in the AS channel, our analysis suggests for instance 
an uncertainty for the mass  as large as 5\%  if only 20 modes are 
considered
 (for the closest models, effective temperatures are compatible within 60K, and radii and luminosities differ by $\sim$ 2\%).
With the method based on $ \delta_{sls}$, 
an uncertainty of 5\% for the mass remains for up to 50 considered modes
(for the closest models, effective temperatures are compatible within 60K, radii differ 
by 3\% and luminosities are equal).


\begin{table}[h]
\caption{Accuracy $\delta M /M$ on mass determination, for a frequency accuracy of 0.2 $\mu$Hz, obtained from different methods of evaluating spectra difference and for different values of considered modes.} 
\begin{tabular}{c|c|c|c} \hline
Nf of modes  & 20 & 40 & 60 \\
\hline
 $\delta M /M$  with $\delta \nu$ & 0.02 & 0.01 & $<$0.01\\
\hline
$\delta M /M$  with $\delta_{sls}$ & $>$ 0.07 & 0.02 & 0.01\\
\hline
\noalign{\smallskip}
\end{tabular}
\label{tabres2}
 \end{table} 

\begin{table}[h]
\caption{Accuracy $\delta L /L$ on luminosity determination, for a frequency accuracy of 0.2 $\mu$Hz, obtained from different methods of evaluating spectra difference and for different values of considered modes.} 
\begin{tabular}{c|c|c|c} \hline
Nf of modes  & 20 & 40 & 60 \\
\hline
 $\delta L /L$  with $\delta \nu$ & 0.06 & 0.04 & 0.02\\
\hline
$\delta L /L$  with $\delta_{sls}$ &  0.1 & 0.05 & 0.05\\
\hline
\noalign{\smallskip}
\end{tabular}
\label{tabres2L}
 \end{table} 

\begin{table}[h]
\caption{Accuracy $\delta R /R$ on radius determination, for a frequency accuracy of 0.2 $\mu$Hz, obtained from different methods of evaluating spectra difference and for different values of considered modes.} 
\begin{tabular}{c|c|c|c} \hline
Nf of modes  & 20 & 40 & 60 \\
\hline
 $\delta R /R$  with $\delta \nu$ & $<$0.01 & $<$0.01 & $<$0.01\\
\hline
$\delta R /R$  with $\delta_{sls}$ & 0.02 & 0.01 & $<$0.01\\
\hline
\noalign{\smallskip}
\end{tabular}
\label{tabres2R}
 \end{table}

\begin{table}[h]
\caption{Same as Table \ref{tabres2} for a frequency accuracy of 0.3 $\mu$Hz}
\begin{tabular}{c|c|c|c} \hline
Nf of modes  & 20 & 40 & 60 \\
\hline
 $\delta M /M$  with $\delta \nu$ & 0.05 & 0.02 & 0.02\\
\hline
 $\delta M /M$  with $\delta_{sls}$ & $>$ 0.07 & 0.06 & 0.04\\
\hline
\noalign{\smallskip}
\end{tabular}
\label{tabres1}
 \end{table}

\begin{table}[h]
\caption{Same as Table \ref{tabres2L} for a frequency accuracy of 0.3 $\mu$Hz}
\begin{tabular}{c|c|c|c} \hline
Nf of modes  & 20 & 40 & 60 \\
\hline
 $\delta L /L$  with $\delta \nu$ & 0.09 & 0.05 & 0.03\\
\hline
 $\delta L /L$  with $\delta_{sls}$ & $>$0.1 & $>$0.1 & $>$0.1\\
\hline
\noalign{\smallskip}
\end{tabular}
\label{tabres1L}
 \end{table} 

\begin{table}[h]
\caption{Same as Table \ref{tabres2R} for a frequency accuracy of 0.3 $\mu$Hz}
\begin{tabular}{c|c|c|c} \hline
Nf of modes  & 20 & 40 & 60 \\
\hline
 $\delta R /R$  with $\delta \nu$ & 0.01 & $<$0.01 & $<$0.01\\
\hline
 $\delta R /R$  with $\delta_{sls}$ & 0.025 & 0.015 & 0.01\\
\hline
\noalign{\smallskip}
\end{tabular}
\label{tabres1R}
 \end{table}



Our work is a first step toward a more systematic study, as 
it only takes into account  variations in
the basic input physics of stellar models such as the mixing length
parameter, the helium abundance and to a certain extent the metallicity.
A more systematic analysis should also explore the
 effects of different convection models, of core
overshooting and of rotation. These uncertainties in current stellar evolution
theory will certainly
increase the degeneracy of the pulsation spectra ({\it i.e} the limit on the mass, luminosity or radius 
mentioned above may be significantly larger).
Moreover, we have not explored the effect of uncertainties in the stability
analysis calculation, such as the sensitivity to boundary conditions. 
The large and small separations are expected to be less sensitive than absolute frequencies to boundary conditions used in the computation. 
Obviously, much additional works need to be done.
Our work, however, provides an idea of the level of accuracy 
 required by asteroseismological studies in order to improve 
 exoplanet-host star
 parameters, crucial information for future projects such as PLATO.


\begin{acknowledgements} 
C. M-M and I. B acknowledge support from Agence Nationale de la Recherche
under the ANR project number NT05-3 42319.
\end{acknowledgements}

\end{document}